\documentclass[12pt]{article}
\usepackage{makeidx}
\usepackage{amsmath}
\usepackage{amsfonts}
\usepackage{amssymb}
\usepackage{graphicx}
\usepackage{cite}
\usepackage{xcolor}
\usepackage{amsthm}

\makeatletter
\g@addto@macro\th@plain{\thm@headpunct{:}}
\makeatother
\theoremstyle{plain}
\newtheorem*{acknowledgement*}{Acknowledgement}
\input epsf.sty
\makeatletter \@addtoreset{equation}{section}
\renewcommand{\theequation}{\thesection.\arabic{equation}}
\input{tcilatex}

\begin{document}

\title{\textbf{On the partial breaking of $\mathcal{N}=2$ rigid supersymmetry with complex hypermultiplet}}

\footnotetext{\footnotesize $^\dagger$Corresponding author, E-mail: mohamedvall.ma@gmail.com}

\author{M.N. El Kinani, M. Vall$^{\dagger}$ \\
{\small Lab of High Energy Physics, Modeling and Simulations, Faculty of Sciences, Rabat, Morocco}}

\maketitle

\begin{abstract}
We study partial supersymmetry breaking in effective $\mathcal{N}=2$ U$ \left( 1\right) ^{n}$ gauge theory coupled to complex hypermultiplets by using the method of \textrm{"{\small arXiv:1501.07842}" to which we refer to as ADFT method}. We derive the generalisation of the symplectic invariant ADFT formula $\zeta _{a}=\frac{1}{2}\varepsilon _{abc}\left(\mathcal{P}^{bM}\mathcal{C}_{MN}\mathcal{P}^{cN}\right) $ capturing information on partial breaking. Our extension of this anomaly is expressed like $d_{a}=\frac{1}{2}\varepsilon _{abc}\mathbb{P}^{bM}\mathcal{C}_{MN}\mathbb{P}^{cN}+\mathcal{J}_{a}$; the generalized moment maps $\mathbb{P}^{aM}$ contain $\mathcal{P}^{aM}$ and depend as well on electric/magnetic coupling charges $G^{M}=\left( \eta ^{i},g_{i}\right) $, the $\mathcal{J}_{a}$ is an extra contribution \textrm{\ induced by Killing isometries in complex hypermatter sector}. Using SP$\left(2n,\mathbb{R}\right)$ symplectic symmetry, we also give the $\mathcal{N}=2$ partial breaking condition and derive the model of "{\small arXiv: 1204.2141}" by a particular realisation of $d_{a}$ anomaly.
\end{abstract}


\section{Introduction}

The partial breaking of rigid $\mathcal{N}=2$ supersymmetric field theory
was widely studied in the literature \textrm{\cite%
{Antoniadis:1995vb,Ferrara:1995xi,Ivanov:1997mt,David:2003dh,Fujiwara:2005hj,Antoniadis:2012cg,Andrianopoli:2015wqa,Andrianopoli:2015rpa}
}. To our knowledge, the first model realizing the $\mathcal{N}=2$ partial
breaking was introduced by \emph{Antoniadis, Partouche and Taylor} (APT)$\ $%
in \textrm{\cite{Antoniadis:1995vb}} where the partial breaking, for an $%
\mathcal{N}=2$ effective pure abelian gauge theory, was interpreted as
resulting from the presence of magnetic Fayet-Ilioupoulos (FI) charges
beside the electric ones. The presence of these FI charges allow to evade
the no-go theorem which forbids such phenomenon to occur \textrm{\cite%
{Cecotti:1984rk,Cecotti:1985mx}}. Recently the interpretation of the APT
model in terms of triplets of symplectic hyper-Kahler moment maps $\left( 
\mathcal{P}^{a}\right) ^{M}\equiv \mathcal{P}^{aM}$ was given by \emph{%
Andrianopoli, D'Auria, Ferrara and Trigiante }(ADFT) in \textrm{\cite%
{Andrianopoli:2015wqa}}. There, the authors showed that the partial breaking
is induced by a non vanishing symplectic invariant isotriplet $\boldsymbol{C}%
_{B}^{A}=\boldsymbol{\zeta }_{c}\left( \tau ^{c}\right) _{B}^{A}$ deforming
the $SU\left( 2\right) _{R}$ invariant APT scalar potential $\delta _{B}^{A}%
\mathcal{V}_{APT}$ in the rigid Ward identity. This $SU\left( 2\right) _{R}$
triplet $\boldsymbol{\zeta }_{c}$ reads in terms of the $\mathcal{P}^{aM}$'s
as follows 
\begin{equation}
\boldsymbol{\zeta }_{c}=\frac{1}{2}\varepsilon _{abc}\mathcal{P}^{aM}%
\mathcal{C}_{MN}\mathcal{P}^{bN}  \label{cab1}
\end{equation}%
where $\tau ^{c}$ stand for the three Pauli matrices, $\varepsilon _{abc}$
the completely antisymmetric tensor in 3D and the $2n\times 2n$ matrix%
\begin{equation}
\mathcal{C}_{MN}=%
\begin{pmatrix}
\mathbf{0}_{n} & \mathbf{I}_{n} \\ 
-\mathbf{I}_{n} & \mathbf{0}_{n}%
\end{pmatrix}
\label{spm}
\end{equation}%
is the $SP\left( 2n,\mathbb{R}\right) $ symplectic metric of the scalar
manifold of the Coulomb branch of the $\mathcal{N}=2$ supersymmetric $%
U\left( 1\right) ^{n}$ effective gauge theory. The $\mathcal{P}^{aM}$ moment
maps are obtained from the gauging \textrm{of Killing vectors of }the
hypermultiplet scalar manifold SO$\left( 1,4\right) /SO\left( 4\right) $ in
the rigid limit of gauged $\mathcal{N}=2$ supergravity \textrm{\cite%
{Andrianopoli:2015rpa,Laamara:2017hdl}}. 
In this rigid limit, the observable sector contains only the $\mathcal{N}=2$ supersymmetric $U\left( 1\right)^{n}$ vector multiplets while gravitation and hypermultiplet are in the
hidden sector; and then the study of the coupling of the APT model, by using
the ADFT method \textrm{\cite{Andrianopoli:2015wqa}, }to observable
hypermatter is still missing.\newline
The aim of the present paper is to fill this gap in the literature by
studying the extension of the $\mathcal{N}=2$ ADFT method to include gauge
invariant couplings with observable complex hypermatter. Here, we use the $%
\mathcal{N}=1$ language to deal with $\mathcal{N}=2$ supermultiplets; the
complex scalars $q^{1},q^{2}$ of a hypermultiplet carry opposite gauge
charges under $U\left( 1\right) ^{n}$ and are thought of as scalars of two $%
\mathcal{N}=1$ multiplets $Q^{u}=\left \{ q^{u},\chi ^{u}\right \} $ with
complex $\chi ^{u}\ $standing for the fermionic partners of the $q^{u}$ in
the complex hypermultiplet. The Weyl fermions $\chi ^{1},\chi ^{2}$ carry
opposite gauge charge under $U\left( 1\right) ^{n}$, the same charges as for 
$q^{1},q^{2}$. By exhibiting the $SP\left( 2n,\mathbb{R}\right) $ symplectic
structure of the extended model, and following the ADFT method, we show in
this study that the partial breaking of effective $\mathcal{N}=2$
supersymmetric $U\left( 1\right) ^{n}$ theory coupled to a complex
hypermultiplet is due to a generalised moment maps $\mathbb{P}^{aM}$
extending the $\mathcal{P}^{aM}$'s of ADFT appearing in eq(\ref{cab1}) and
having the form $\mathbb{P}^{aM}=\mathcal{P}^{aM}+\delta \mathcal{P}^{aM}$.
The moment map deformation $\delta \mathcal{P}^{aM}$ is induced by the
coupling of APT theory to the complex hypermultiplet and so depends on the
symplectic gauge coupling constants $\left( g_{i},\eta ^{i}\right) $ and
also on $\left \langle q^{A}\right \rangle $, the VEVs of the complex fields.
The above ADFT deformation $\zeta _{c}$ extends, in presence of observable
complex hypermatter, as follows 
\begin{equation}
d_{a}=\frac{1}{2}\varepsilon _{abc}\mathbb{P}^{bM}\mathcal{C}_{MN}\mathbb{P}%
^{cN}+\frac{1}{2}\Omega _{uv}\mathcal{N}^{u}\tau _{a}\mathcal{\bar{N}}^{v}
\end{equation}%
and is given by the sum of four contributions like $d_{a}=\zeta _{a}+\alpha
_{a}+\beta _{a}+\mathcal{J}_{a}$ where $\zeta _{a}$ is as in eq(\ref{cab1})
and the others coming from hypermatter and its couplings with gauge degrees
of freedom. In this relation, the first three contributions $\zeta
_{a}+\alpha _{a}+\beta _{a}$ are proportional to the $SP\left( 2n,\mathbb{R}%
\right) $ metric $\mathcal{C}_{MN}$ of the Coulomb branch, while $\mathcal{J}%
_{a}$ is proportional to the symplectic $\Omega _{uv}$ in the hypermatter
sector. After giving the $\mathcal{N}=2$ supersymmetry partial breaking
condition in terms of above abstract isovector $d_{a}$,\ we turn to show
that the \emph{Antoniadis, Derendinger and Jacot\ }(ADJ) scalar potential $%
\mathcal{V}_{ADJ}$ and the partial supersymmetry breaking condition obtained
in \textrm{\cite{Antoniadis:2012cg}} can be recovered by making a particular
choice of the components of $\mathbb{P}^{aM}$ and $\mathcal{N}_{A}^{u}$. 

The organization of this paper is as follows: In\textrm{\ section 2,} we
give a short review of the ADFT method by focussing on key points; in
particular on the derivation of (\ref{cab1}). In\textrm{\ section 3,} we
couple the APT model to a complex hypermultiplet by using the to ADFT method
and develop the study of rigid Ward identity as well as its implication on
the structure of the induced scalar potential and the isovector anomaly%
\textrm{.} In\textrm{\ section 4}, we first give the condition for $\mathcal{%
N}=2$ supersymmetry partial breaking to occur; then we re-derive the ADJ
model of \textrm{\cite{Antoniadis:2012cg}} by choosing particular values of
the components of the moment maps $\mathbb{P}^{aM}$. \textrm{Section 5} is
devoted to the conclusion; 

\section{Rigid Ward identity in ADFT method}

In this section, we give the main lines of the derivation of the scalar
potential $\mathcal{V}_{\text{APT}}^{\mathcal{N}=2}$ and the $2\times 2$
matrix anomaly $\boldsymbol{C}_{B}^{A}$ in the rigid limit of $\mathcal{N}=2$
supergarvity coupled to n abelian vector multiplets and one real
hypermultiplet with $SO\left( 4,1\right) /SO\left( 4\right) $ quaternionic
Kahler geometry by following the ADFT method. In gauged $\mathcal{N}=2$
supergravity theory, the induced scalar potential $\mathcal{V}_{\text{sugra}%
} $ is related to the supersymmetric transformations $\delta \psi ^{\tau A}$
of the fermions via the following Ward identity \textrm{\cite%
{Andrianopoli:2015wqa}},%
\begin{equation}
\delta _{B}^{A}\mathcal{V}_{\text{sugra}}=\dsum \limits_{\tau }\alpha _{\tau
}\delta _{C}\psi ^{\tau A}\bar{\delta}^{C}\bar{\psi}_{B}^{\tau }  \label{wid}
\end{equation}%
Here, the summation is taken over all the $\psi ^{\tau A}$ fermions in the
gauged $\mathcal{N}=2$ supergravity theory; it includes the two gravitini,
the gaugini and the hyperini. The $\alpha _{\tau }$ constants are positive
for the spin $\frac{1}{2}$\textrm{\ }fermions and negative for the gravitini.%
\newline
In the rigid limit considered in \textrm{\cite%
{Andrianopoli:2015rpa,Laamara:2017hdl}}, 
the above supergravity Ward
identity (\ref{wid}) becomes%
\begin{equation}
\mathcal{G}_{i\bar{j}}\left( W^{i}\right) _{C}^{A}(\bar{W}^{\bar{j}%
})_{B}^{C}=\delta _{B}^{A}\mathcal{V}_{\text{APT}}^{\mathcal{N}=2}+%
\boldsymbol{C}_{B}^{A}  \label{gwi}
\end{equation}%
where $\mathcal{V}_{\text{APT}}^{\mathcal{N}=2}$ is the APT potential and $%
\boldsymbol{C}_{B}^{A}=\boldsymbol{\zeta }_{c}\left( \tau ^{c}\right)
_{B}^{A}$ \textrm{is an anomalous term} which has been shown to have an
interpretation in terms of symplectic hyper-Kahler moment maps $\mathcal{P}%
^{aM}$ as in (\ref{cab1}). The other quantities involved in this identity
are briefly described below. First, $\mathcal{G}_{i\bar{j}}$ is the metric
of the rigid special Kahler manifold of the Coulomb branch, it is given by $%
\func{Im}\mathcal{F}_{ij}$ which is the imaginary part of the second
derivative of the holomorphic prepotential $\mathcal{F}$ of the effective $%
\mathcal{N}=2$ supersymmetric U$\left( 1\right) ^{n}$ gauge theory \textrm{%
\cite{Antoniadis:2012cg}}. The $n$ quantities $\left( W^{i}\right) _{B}^{A}$
are 2$\times $2 matrices involved in the supersymmetric transformations of
the $n$ gaugini doublets $\lambda ^{iA}$; they are given by 
\begin{equation}
\delta _{B}\lambda ^{iA}=\left( W^{i}\right) _{B}^{A}  \label{trs}
\end{equation}%
where $\left( W^{j}\right) _{B}^{A}$ matrices read explicitly as follows 
\begin{equation}
\left( W^{j}\right) _{B}^{A}=%
\begin{pmatrix}
iD^{j} & -\sqrt{2}F^{j} \\ 
\sqrt{2}F^{j}+\frac{1}{2}\sqrt{2}m^{j} & -iD^{j}%
\end{pmatrix}
\label{wjab}
\end{equation}%
whith $\frac{1}{\sqrt{2}}m^{j}$ has an interpretation in terms of \emph{%
magnetic} FI charges \textrm{\cite{Antoniadis:2012cg}}.
By substituting the auxiliary fields $\left( F^{j},\bar{F}%
^{j},D^{j}\right) $ by their explicit expressions derived from the effective
field action of the $\mathcal{N}=2$ supersymmetric gauge theory, 
we can bring the above matrices
into the following form 
\begin{equation}
\left( W^{i}\right) _{B}^{A}=i\mathcal{G}^{i\bar{j}}\left( \tau _{a}\right)
_{B}^{A}W^{ai}\qquad ,\qquad W^{ai}=\mathcal{P}_{M}^{a}\bar{U}_{\bar{j}}^{M}
\label{wab}
\end{equation}%
where $\mathcal{P}_{M}^{a}$ are the moment maps of the 4-dim hyperKahler
geometry in the hidden sector of the rigid limit of $\mathcal{N}=2$ gauged
supergravity \cite{Andrianopoli:2015rpa,Laamara:2017hdl}. 
An explicit
expression of these $\mathcal{P}_{M}^{a}$'s is given by the values of APT
model; its reads in terms of $\mathcal{N}=2$ electric $e_{i}^{a}$ and
magnetic $m^{ai}$ Fayet-Iliopoulos charges as follows%
\begin{equation}
\mathcal{P}^{aM}=\sqrt{2}\left( 
\begin{array}{c}
m^{ai} \\ 
e_{i}^{a}%
\end{array}%
\right)  \label{pm}
\end{equation}%
In this formulation, the electric $e_{i}^{a}$'s and the magnetic $m^{ai}$'s
are thought of as real isotriplets; and so the moment maps are also real
isotriplets carrying moreover a symplectic quantum number. The complex
quantity $U_{i}^{M}$ is the gradient of the holomorphic SP$\left( 2n,\mathbb{%
R}\right) $ symplectic section $V^{M}$ of the rigid special Kahler geometry.
In the local coordinate frame where the homogeneous $\left(
X^{0},X^{I}\right) $ are taken as $\left( 1,\delta _{i}^{I}z^{i}\right) $,
the expressions of $V^{M}$ and $U_{i}^{M}$ read like%
\begin{equation}
V^{M}=\left( 
\begin{array}{c}
z^{i} \\ 
\mathcal{F}_{i}%
\end{array}%
\right) \quad ,\quad U_{i}^{M}=\frac{\partial }{\partial z^{i}}V^{M}
\label{rigs}
\end{equation}%
The complex $z^{i}$ are the n scalar fields belonging to the $\mathcal{N}=2$
supersymmetric U$\left( 1\right) ^{n}$ vector supermultiplets and $\mathcal{F%
}_{i}$ are the symplectic dual of $z^{i}$; they are given by the gradient of
the holomorphic prepotential $\mathcal{F}\left( z\right) $ of the effective
theory. Putting (\ref{wab}) back into (\ref{gwi}), we can obtain explicit
expressions of the APT potential $\mathcal{V}_{\text{APT}}^{\mathcal{N}=2}$
and the rigid anomaly $\boldsymbol{C}_{B}^{A}$ in terms of the following
quantities: $\left( i\right) $ the symplectic isotriplet moment maps $%
\mathcal{P}_{M}^{a}$ of the $SO\left( 4,1\right) /SO\left( 4\right) $
quaternionic Kahler manifold, $\left( ii\right) $ the metric $\mathcal{G}_{i%
\bar{j}}$ of the special Kahler manifold; and $\left( iii\right) $ the
holomorphic sections $U_{i}^{M}$; they read as follows 
\begin{equation}
\begin{tabular}{lll}
$\mathcal{V}_{\text{APT}}^{\mathcal{N}=2}$ & $=$ & $\frac{1}{2}\mathcal{P}%
_{M}^{a}\mathcal{M}^{MN}\mathcal{P}_{N}^{a}$ \\ 
$\boldsymbol{C}_{B}^{A}$ & $=$ & $\frac{1}{2}\varepsilon _{abc}\left( 
\mathcal{P}_{M}^{a}\mathcal{C}^{MN}\mathcal{P}_{N}^{b}\right) \left( \tau
^{c}\right) _{B}^{A}$%
\end{tabular}%
\end{equation}%
where $\mathcal{C}^{MN}$ is the invariant symplectic metric of SP$\left( 2n,%
\mathbb{R}\right) $ and the symplectic coupling matrix $\mathcal{M}^{MN}$ is
given by 
\begin{equation}
\mathcal{M}^{MN}=2U_{i}^{M}\mathcal{G}^{i\bar{j}}\bar{U}_{\bar{j}}^{N}+i%
\mathcal{C}^{MN}
\end{equation}%
It is related to the prepotential $\mathcal{F}$ as follows \textrm{\cite%
{Andrianopoli:2006ub,Andrianopoli:2015wqa}}%
\begin{equation}
\mathcal{M}_{MN}=%
\begin{pmatrix}
\func{Im}\mathcal{F}+\func{Re}\mathcal{F}\left( \func{Im}\mathcal{F}\right)
^{-1}\func{Re}\mathcal{F} & -\func{Re}\mathcal{F}\left( \func{Im}\mathcal{F}%
\right) ^{-1} \\ 
-\left( \func{Im}\mathcal{F}\right) ^{-1}\func{Re}\mathcal{F} & \left( \func{%
Im}\mathcal{F}\right) ^{-1}%
\end{pmatrix}
\label{mat}
\end{equation}%
Notice that in the left hand of the rigid Ward identity (\ref{gwi}), there
is only a contribution coming from the gauge sector of the theory. In next
section, we study the generalisation of this Ward identity by implementing
the contribution coming from hypermatter sector. This requires extending the
APT model by implementing gauge invariant couplings to complex hypermatter.

\section{Coupling APT model to hypermatter}

\label{sec2}

In this section, we use the ADFT method introduced above to study the extension of the APT model by implementing gauge invariant couplings between the $n$ vector multiplets
and $n_{H}$ complex hypermultiplets. Here, we will assume that the complex
scalar fields of the hypermultiplets parameterise the flat complex
hyperKahler manifold $\mathbb{C}^{2n_{H}}$. 

\subsection{Rigid Ward identity in extended ADFT}

The rigid Ward identity of ADFT method coupled to complex hypermatter has two
main contributions: a gauge contribution coming from the pure ADFT sector
and a matter one coming from complex hypermatter. Before studying these
contributions it is interesting to introduce the bosonic and fermionic
degrees of freedom of a complex on shell hypermultiplet $\boldsymbol{H}$.
The usual four real bosonic degrees are described here by two complex fields 
$q_{1},$ $q_{2}$ carrying opposite gauge charges under U$\left( 1\right)
^{n} $ gauge symmetry group of the APT theory. They will be collectively
denoted like $q^{u}$ with $u=1,2$. The fermionic degrees of the complex
hypermultiplet are given by a pair of complex Weyl $\psi $ and $\xi $
carrying opposite U$\left( 1\right) $ gauge charges and are collectively
denoted by $\chi ^{u}=\left( \psi ,\xi \right) $. Bosonic and fermionic
degrees form two chiral supermultiplets $\left \{ q^{u},\chi ^{u}\right \} $
and can be described by $\mathcal{N}=1$ chiral superfields $Q^{u}$ carrying
opposite gauge charges. 
In the case where there are $n_{H}$ complex hypermultiplets $\boldsymbol{H}^{I}$, the bosonic and
fermionic degrees of freedom are described by adding the extra index $%
I=1,...,n_{H}$ \textrm{\cite{Saidi:2008au,Sahraoui:1999je}}. For
convenience, we shall restrict our analysis below to one complex
hypermultiplet by dropping out the $I$\ index; the generalisation for the
particular $\mathbb{C}^{2n_{H}}$ hyperKahler geometry is straightforward and
omitted. In the case $n_{H}=1$, the rigid fermionic transformation
generalising (\ref{trs}) are given by \textrm{\cite{Antoniadis:2012cg}}%
\begin{equation}
\begin{tabular}{lll}
$\delta \lambda ^{iA}$ & $=$ & $\left( \mathcal{W}^{i}\right)
_{B}^{A}\epsilon ^{B}$ \\ 
$\delta \chi ^{u}$ & $=$ & $2\left( \mathcal{N}_{A}^{u}\right) \epsilon ^{A}$%
\end{tabular}
\label{fft}
\end{equation}%
where the $\lambda ^{iA}$'s are the $n$ gaugini doublets of section 2 and
the $\chi ^{u}$ referring to the $2$ hyperini of hypermatter. 
The scalar matrices $\mathcal{W}^{i}$ and $\mathcal{N}_{A}^{u}$ are given by: 
\begin{equation}
\begin{tabular}{lll}
$\mathcal{N}_{A}^{u}$ & $=$ & $K_{A}^{u}G_{M}\bar{V}^{M}$ \\ 
$\left( \mathcal{W}^{i}\right) _{B}^{A}$ & $=$ & $i\mathcal{G}^{i\bar{j}%
}\left( \tau _{a}\right) _{B}^{A}(\mathbb{P}_{M}^{a}\bar{U}_{\bar{j}}^{M})$%
\end{tabular}
\label{k1}
\end{equation}%
with $\bar{V}^{M}$ is the antiholomorphic section of the rigid special
geometry given by (\ref{rigs}). The $G_{M}$ and $K^{Au}$ \textrm{stand
respectively} for the electric- magnetic gauge charges and Killings%
\begin{equation}
\begin{tabular}{lllllll}
$G^{M}$ & $=$ & $\left( 
\begin{array}{c}
\eta ^{i} \\ 
g_{i}%
\end{array}%
\right) $ & $\qquad \text{,}\qquad $ & $G_{M}$ & $=$ & $\left( 
\begin{array}{c}
-g_{i} \\ 
\eta ^{i}%
\end{array}%
\right) $ \\ 
$K^{Au}$ & $=$ & $\left( 
\begin{array}{c}
k^{u} \\ 
\bar{k}_{u}%
\end{array}%
\right) $ & $\qquad \text{,}\qquad $ & $K_{A}^{u}$ & $=$ & $\left( 
\begin{array}{c}
-\bar{k}_{u} \\ 
k^{u}%
\end{array}%
\right) $%
\end{tabular}
\label{gn}
\end{equation}%
with $k^{u}$ giving the Killing vectors of the complex hypermultiplet
manifold explicitly given by%
\begin{equation}
k^{u}=-i%
\begin{pmatrix}
q^{1} \\ 
-q^{2}%
\end{pmatrix}%
\qquad \text{,}\qquad \bar{k}_{u}=i%
\begin{pmatrix}
\bar{q}_{1} \\ 
-\bar{q}_{2}%
\end{pmatrix}
\label{xkv}
\end{equation}%
In the second relation of the system of equations (\ref{k1}), the $\mathcal{G%
}^{i\bar{j}}$ and\ $\bar{U}_{\bar{j}}^{M}$ are as in (\ref{wab}); but the $%
\mathbb{P}_{M}^{a}$'s are generalized moment maps. 
Recall that the real scalars ($\varphi ,\phi^{a}$) parametrize the real hyper-Kahler manifold $\frac{SO\left( 4,1\right) }{SO\left( 4\right) }$ which lives in the hidden sector of the rigid limit of $\mathcal{N}=2$ gauged supergravity \textrm{\cite{Andrianopoli:2015rpa,Laamara:2017hdl}}; 
while the complex scalars $q^{u}=\left(q^{1},q^{2}\right)$ are coordinates of 
the hyperKahler manifold $\mathbb{C}^{2}$. 
Thus, one can expect that the $\mathbb{P}_{M}^{a}$ has two main
contributions; the previous $\mathcal{P}_{M}^{a}$ coming from the $\frac{%
SO\left( 4,1\right) }{SO\left( 4\right) }$ factor of the hidden sector and
an extra $\mathcal{R}_{M}^{a}$ descending from the complex hypermultiplet
parameterising $\mathbb{C}^{2}$. So, we can define $\mathbb{P}_{M}^{a}$ as
follows%
\begin{equation}
\mathbb{P}_{M}^{a}=\mathcal{P}_{M}^{a}+\mathcal{R}_{M}^{a}\qquad ,\qquad
\left. \mathbb{P}^{aM}\right \vert _{ADFT}=\mathcal{P}^{aM}  \label{mt}
\end{equation}%
with%
\begin{equation}
\mathcal{P}^{aM}=\sqrt{2}\left( 
\begin{array}{c}
m^{ai} \\ 
e_{i}^{a}%
\end{array}%
\right) \qquad ,\qquad \mathcal{R}^{aM}=\sqrt{2}\left( 
\begin{array}{c}
\mathcal{M}^{ai} \\ 
\mathcal{E}_{i}^{a}%
\end{array}%
\right)  \label{rm}
\end{equation}%
where $\mathcal{R}^{aM}$ is due to hypermatter. The splitting (\ref{mt})
implies that $\left( \mathcal{W}^{i}\right) _{B}^{A}$ splits as well as the
sum of two terms as follows%
\begin{equation}
\left( \mathcal{W}^{i}\right) _{B}^{A}=\left( W^{i}\right) _{B}^{A}+\left(
Y^{i}\right) _{B}^{A}
\end{equation}%
with $\left( W^{i}\right) _{B}^{A}$ given by (\ref{wab}) of the ADFT
and 
\begin{equation}
\left( Y^{i}\right) _{B}^{A}=i\mathcal{G}^{i\bar{j}}\left( \tau _{a}\right)
_{B}^{A}(\mathcal{R}_{M}^{a}\bar{U}_{\bar{j}}^{M})
\end{equation}%
Moreover, because of the presence of complex hypermatter, the rigid Ward
identity (\ref{gwi}) extends as follows%
\begin{equation}
\mathcal{G}_{i\bar{j}}\left( \mathcal{W}^{i}\right) _{C}^{A}(\mathcal{\bar{W}%
}^{\bar{j}})_{B}^{C}+2\left( \mathcal{N}_{B}^{u}\right) \left( \mathcal{\bar{%
N}}_{u}^{A}\right) =\delta _{B}^{A}\mathcal{V}_{\text{scal}}^{\mathcal{N}=2}+%
\boldsymbol{D}_{B}^{A}  \label{rwi}
\end{equation}%
where now $\mathcal{V}_{\text{scal}}^{\mathcal{N}=2}$ is the scalar
potential of the deformed APT model which will be shown later on to lead to
the ADJ potential $\mathcal{V}_{\text{ADJ}}^{\mathcal{N}=2}$ and \textrm{a
variant of it involving dyonic gauge couplings}. The $\boldsymbol{D}_{B}^{A}$
is an anomalous term that will be determined as well later on. These two $%
\mathcal{V}_{\text{scal}}^{\mathcal{N}=2}$ and $\boldsymbol{D}_{B}^{A}$
quantities are functions of the degrees of freedom of the complex
hypermultiplets; in particular of the fields $\left( q^{1},q^{2}\right) $;
they reduce to $\mathcal{V}_{\text{APT}}^{\mathcal{N}=2}$ and $\boldsymbol{C}%
_{B}^{A}$ of the ADFT if the couplings to complex hypermultiplet are
turned off. The left hand side of the rigid Ward identity (\ref{rwi}) has
now two blocks $\boldsymbol{G}_{A}^{B}$ and $\boldsymbol{M}_{A}^{B}$ given by%
\begin{equation}
\begin{tabular}{lll}
$\boldsymbol{G}_{B}^{A}$ & $=$ & $\mathcal{G}_{i\bar{j}}\left( \mathcal{W}%
^{i}\right) _{C}^{A}(\mathcal{\bar{W}}^{\bar{j}})_{B}^{C}$ \\ 
$\boldsymbol{M}_{B}^{A}$ & $=$ & $2\left( \mathcal{N}_{B}^{u}\right) \left( 
\mathcal{\bar{N}}_{u}^{A}\right) $%
\end{tabular}
\label{sp}
\end{equation}%
with fermionic field shifts $\left( \mathcal{W}^{i}\right) _{B}^{A}$ and $%
\mathcal{N}_{A}^{u}$ as in (\ref{k1}). In the next subsection we compute the
above $\boldsymbol{G}_{B}^{A}$ and $\boldsymbol{M}_{B}^{A}$ in term of the
moment maps (\ref{mt}) and compare the obtained expressions with the ones
given by ADFT method.

\subsection{Computing eqs(\protect \ref{k1})}

First, we calculate the gauge sector contribution to the rigid Ward identity
given by $\boldsymbol{G_{B}^{A}}$; and turn after to the contribution $%
\boldsymbol{M}_{B}^{A}$ coming from complex hypermatter.

\subsubsection{APT sector}

The explicit expression of the gauge sector contribution $\boldsymbol{%
G_{B}^{A}}=\mathcal{G}_{i\bar{j}}\left( \mathcal{W}^{i}\right) _{C}^{A}(%
\mathcal{\bar{W}}^{\bar{j}})_{B}^{C}$ to the rigid Ward identity (\ref{rwi})
is obtained by substituting $\left( \mathcal{W}^{i}\right) _{C}^{A}$ by its
expression in (\ref{k1}). We find after rearranging terms the following
expression%
\begin{equation}
\boldsymbol{G}_{B}^{A}=\frac{1}{2}\delta _{B}^{A}\mathbb{P}_{M}^{a}\mathcal{M%
}^{MN}\mathbb{P}_{N}^{a}+\frac{1}{2}\varepsilon _{abc}\left( \mathbb{P}%
_{M}^{a}\mathcal{C}^{MN}\mathbb{P}_{N}^{b}\right) \left( \tau ^{c}\right)
_{B}^{A}  \label{ga}
\end{equation}%
This 2$\times $2 matrix has four terms that can be viewed as the sum of two
SU$\left( 2\right) _{R}$ blocks like $\delta _{B}^{A}\mathcal{S}_{0}+%
\mathcal{S}_{B}^{A}$ with $Tr\mathcal{S}_{B}^{A}=0$. These two blocks are
precisely the contributions to the scalar potential $\mathcal{V}_{\text{scal}%
}^{\mathcal{N}=2}$ and the $\boldsymbol{D}_{B}^{A}$ anomaly coming from the
transformations of gaugini.

$\bullet $ \emph{Diagonal }$\mathcal{S}_{0}$\emph{\ term }\newline
The isosinglet $\mathcal{S}_{0}$ is given by $\frac{1}{2}Tr\left( 
\boldsymbol{G}_{B}^{A}\right) $; it reads in terms of the generalised moment
maps $\mathbb{P}_{M}^{a}$ as follows%
\begin{equation}
\mathcal{S}_{0}=\frac{1}{2}\mathbb{P}_{M}^{a}\mathcal{M}^{MN}\mathbb{P}%
_{N}^{a}  \label{s0}
\end{equation}%
where $\mathcal{M}^{MN}$ is the inverse of (\ref{mat}) and where the
generalised moment maps $\mathbb{P}_{M}^{a}$ is given by (\ref{mt}); it has
an extra $\mathcal{R}_{M}^{a}$ contribution coming from the gauge invariant
complex hypermatter couplings. By substituting $\mathbb{P}_{M}^{a}$ in (\ref%
{s0}) by $\mathcal{P}_{M}^{a}+\mathcal{R}_{M}^{a}$, the above $\mathcal{S}%
_{0}$ splits as the sum of three terms as follows%
\begin{equation}
\mathcal{S}_{0}=\frac{1}{2}\mathcal{P}_{M}^{a}\mathcal{M}^{MN}\mathcal{P}%
_{N}^{a}+\frac{1}{2}\mathcal{R}_{M}^{a}\mathcal{M}^{MN}\mathcal{R}_{N}^{a}+%
\mathcal{P}_{M}^{a}\mathcal{M}^{MN}\mathcal{R}_{N}^{a}  \label{ss}
\end{equation}%
where the two first terms depend on $\mathcal{P}_{M}^{a}$; but the last
contribution is independent from the FI charges. So for the limit where the
complex hypermatter couplings to gauge sector are turned off; i.e: the limit 
$\mathcal{R}_{M}^{a}\rightarrow 0$, the isosinglet $\mathcal{S}_{0}$ reduces
to the ADFT relation and so the scalar potential coincides with the one of
the APT\ model which in ADFT method is nothing but a particular field
realisation of the following expression 
\begin{eqnarray}
\mathcal{S}_{0}^{\left( ADFT\right) } &=&\frac{1}{2}\mathcal{P}_{M}^{a}%
\mathcal{M}^{MN}\mathcal{P}_{N}^{a}  \notag \\
&=&\mathcal{G}^{i\bar{j}}\left[ e_{i}^{a}-\mu ^{ka}\mathcal{F}_{ki}\right] 
\overline{\left[ e_{j}^{a}-\mu ^{ka}\mathcal{F}_{kj}\right] }
\end{eqnarray}%
expressed in terms of the electric $e_{i}^{a}$ and magnetic $\mu ^{ka}$ FI
charges. In the limit $\mathcal{P}_{M}^{a}\rightarrow 0$, describing the
case where electric and magnetic FI charges are turned off, we have 
\begin{eqnarray}
\left. \mathcal{S}_{0}\right \vert _{\mathcal{P}_{M}^{a}\rightarrow 0} &=&%
\frac{1}{2}\mathcal{R}_{M}^{a}\mathcal{M}^{MN}\mathcal{R}_{N}^{a}
\label{adm} \\
&=&\mathcal{G}^{i\bar{j}}\left[ \mathcal{E}_{i}^{a}-\mathcal{M}^{ka}\left( 
\mathcal{F}\right) _{ki}\right] \overline{\left[ \mathcal{E}_{j}^{a}-%
\mathcal{M}^{ka}\left( \mathcal{F}\right) _{kj}\right] }  \notag
\end{eqnarray}%
where the the electric $\mathcal{E}_{i}^{a}$ and magnetic $\mathcal{M}^{ka}$
are field dependent couplings. They are functions of the complex matter
scalars $q_{u}$ and the electric $g_{i}$ and magnetic $\eta ^{i}$ gauge
coupling charges. If using the particular realisation given by eq(\ref{ap})
namely%
\begin{equation}
\mathcal{E}_{i}^{a}=g_{i}\digamma ^{a}\qquad \text{,}\qquad \mathcal{M}%
^{ia}=\eta ^{i}\digamma ^{a}
\end{equation}%
these two factorised relations can be combined into a symplectic object as
follows%
\begin{equation}
\mathcal{R}_{M}^{a}=G_{M}\digamma ^{a}  \label{rg}
\end{equation}%
where $\mathcal{R}_{M}^{a}$ is as in (\ref{gn}). So, we have%
\begin{eqnarray}
\left. \mathcal{S}_{0}\right \vert _{\mathcal{P}_{M}^{a}\rightarrow 0} &=&%
\frac{1}{2}G_{M}\mathcal{M}^{MN}G_{N}\times \left \vert \digamma ^{a}\right
\vert ^{2}  \notag \\
&=&\mathcal{G}^{i\bar{j}}\left[ g_{i}-\eta ^{l}\mathcal{F}_{li}\right] 
\overline{\left[ g_{j}-\eta ^{l}\mathcal{F}_{lj}\right] }\times \left \vert
\digamma ^{a}\right \vert ^{2}
\end{eqnarray}%
By using (\ref{rg}), the third cross term in (\ref{ss}) reads as follows%
\begin{equation}
\mathcal{P}_{M}^{a}\mathcal{M}^{MN}\mathcal{R}_{N}^{a}=G_{M}\mathcal{M}^{MN}%
\mathcal{P}_{N}^{a}\digamma ^{a}
\end{equation}%
It depends on the product of the $\mathcal{P}_{N}^{a}$ Fayet-Iliopoulos and
the $G_{M}$ gauge electric- magnetic coupling charges. The general
expression of $\mathcal{S}_{0}$ with non vanishing $\mathcal{P}_{M}^{a}$ and 
$\mathcal{R}_{M}^{a}$ gives the contribution to the scalar potential $%
\mathcal{V}_{\text{ADJ}}^{\mathcal{N}=2}$ in eq(\ref{rwi}) coming from the
gaugino sector.

$\bullet $ \emph{Anomalous term }$\mathcal{S}_{B}^{A}$\newline
The traceless matrix $\mathcal{S}_{B}^{A}$ contributes to the $\boldsymbol{D}%
_{B}^{A}$ anomaly in (\ref{rwi}); it can be presented as an isotriplet that
reads as 
\begin{equation}
\mathcal{S}_{B}^{A}=\dsum \limits_{c=1}^{3}\xi _{c}\left( \tau ^{c}\right)
_{B}^{A}  \label{bc}
\end{equation}%
with%
\begin{equation}
\xi _{c}=\frac{1}{2}\varepsilon _{abc}\mathbb{P}_{M}^{a}\mathcal{C}^{MN}%
\mathbb{P}_{N}^{b}  \label{cab}
\end{equation}%
where $\mathcal{C}^{MN}$ is the $SP\left( 2n,\mathbb{R}\right) $ symplectic
metric. A non vanishing value of $\xi _{c}$ breaks partially $\mathcal{N}=2$
supersymmetry; it is then interesting to study when the norm $\left \Vert 
\boldsymbol{\xi }\right \Vert $ is different from zero. By substituting $%
\mathbb{P}_{M}^{a}=\mathcal{P}_{M}^{a}+\mathcal{R}_{M}^{a}$ in (\ref{cab}),
we learn that the isovector $\xi _{c}$ splits like the sum of three
isovectors as follows%
\begin{equation}
\xi _{c}=\zeta _{c}+\alpha _{c}+\beta _{c}  \label{abc}
\end{equation}%
with $\zeta _{c}$ the same isotriplet as in the ADFT; but the two
extra $\boldsymbol{\alpha }_{c}$ and $\boldsymbol{\beta }_{c}$ are new
isotriplets induced by the presence of complex hypermatter. These three
isotriplets are given by 
\begin{equation}
\begin{tabular}{lll}
$\zeta _{c}$ & $=$ & $\frac{1}{2}\varepsilon _{abc}\mathcal{P}_{M}^{a}%
\mathcal{C}^{MN}\mathcal{P}_{N}^{b}$ \\ 
$\beta _{c}$ & $=$ & $\frac{1}{2}\varepsilon _{abc}\mathcal{R}_{M}^{a}%
\mathcal{C}^{MN}\mathcal{R}_{N}^{b}$ \\ 
$\alpha _{c}$ & $=$ & $\varepsilon _{abc}\mathcal{P}_{M}^{a}\mathcal{C}^{MN}%
\mathcal{R}_{N}^{b}$%
\end{tabular}
\label{ab}
\end{equation}%
The real isovector $\zeta _{c}$ is due to the electric $e_{i}^{a}$ and the
magnetic $m^{ai}$ Fayet-Iliopoulos charges given by (\ref{pm}). It reads
explicitly as follows 
\begin{equation}
\zeta _{c}=\varepsilon _{abc}\left( m^{ai}e_{i}^{b}-e_{i}^{a}m^{bi}\right)
=2\varepsilon _{abc}m^{ai}e_{i}^{b}
\end{equation}%
A non vanishing of the above $\zeta _{c}$'s, which reads in 3- dim vector
notation like $\vec{e}_{i}\wedge \vec{m}^{i}$, requires at least the non
vanishing of some of the $e_{i}^{a}$ components and some of the $m^{ai}$'s.
However, in absence of magnetic $m^{ai}$ FI, the isotriplet $\zeta _{c}$
vanishes identically. The term $\beta _{c}$ is different from $\zeta _{c}$
as it is induced from another source, it is due to the electric $g_{i}$ and
magnetic $\eta ^{i}$ gauge coupling constants. Using (\ref{rm}), we first
have 
\begin{equation}
\beta _{c}=\varepsilon _{abc}\left( \mathcal{M}^{ai}\mathcal{E}_{i}^{b}-%
\mathcal{E}_{i}^{a}\mathcal{M}^{bi}\right) =2\varepsilon _{abc}\mathcal{M}%
^{ai}\mathcal{E}_{i}^{b}
\end{equation}%
By following the analysis on Dyonic gauge coupling to complex hypermatter
reported in the appendix, we find that electric term $\mathcal{E}_{i}^{a}$
and the magnetic $\mathcal{M}^{ai}$ factorise as follows 
\begin{equation}
\mathcal{E}_{i}^{a}=g_{i}\mathcal{E}^{a}\qquad ,\qquad \mathcal{M}^{ai}=\eta
^{i}\mathcal{M}^{a}  \label{fa}
\end{equation}%
where $\mathcal{E}^{a}$ and $\mathcal{M}^{a}$ triplets with components
depending on the scalar fields of the complex matter hypermultiplet. So, the
above $\beta _{c}$ anomaly becomes 
\begin{equation}
\beta _{c}=2\left( g_{i}\eta ^{i}\right) \varepsilon _{abc}\mathcal{M}^{a}%
\mathcal{E}^{b}
\end{equation}%
Then, non vanishing $\beta _{c}$ requires non vanishing $g_{i}\eta ^{i}$ and
non vanishing $\mathcal{\vec{E}}\wedge \mathcal{\vec{M}}$. However, from the
particular realisation eq(\ref{ap}) we learn that $\mathcal{E}^{a}=\mathcal{M%
}^{a}=\digamma ^{a}$; then $\beta _{c}$ vanishes identically.%
\begin{equation}
\beta _{c}=0
\end{equation}%
Concerning the anomaly $\alpha _{c}$, it is due to both the $\left(
e_{i}^{a},m^{ai}\right) $ electric- magnetic FI and the $\left( g_{i},\eta
^{i}\right) $ electric-magnetic gauge charges. By using (\ref{rm}) and (\ref%
{fa}) as well as setting 
\begin{equation}
g_{i}m^{ai}=\kappa ^{a}\qquad ,\qquad \eta ^{i}e_{i}^{a}=\pi ^{a}
\end{equation}%
we can put the $\alpha _{c}$ triplet into the form 
\begin{equation}
\alpha _{c}=\varepsilon _{abc}\left[ \kappa ^{a}\mathcal{E}^{b}-\pi ^{a}%
\mathcal{M}^{b}\right]
\end{equation}%
Substituting $\mathcal{E}^{a}=\mathcal{M}^{a}=\digamma ^{a}$, we end with%
\begin{equation}
\alpha _{c}=\varepsilon _{abc}\left( \kappa ^{a}-\pi ^{a}\right) \digamma
^{b}
\end{equation}

\subsubsection{More on scalar potential and anomaly $\boldsymbol{D}_{B}^{A}$}

The contribution $\boldsymbol{M}_{B}^{A}=2\mathcal{N}_{B}^{u}\mathcal{\bar{N}%
}_{u}^{A}$ of the hypermatter sector to the rigid Ward identity (\ref{rwi})
comes from the coupling between the $n$ vector $\boldsymbol{V}_{i}^{\mathcal{%
N}=2}$ multiplets and the matter hypermultiplet $\boldsymbol{H}^{\mathcal{N}%
=2}$. Expanding this matrix $\boldsymbol{M}_{B}^{A}$ on Pauli matrices basis
like $\mathcal{J}_{0}\delta _{B}^{A}+\mathcal{J}_{a}\left( \tau ^{a}\right)
_{B}^{A}$ and equating with $2\mathcal{N}_{B}^{u}\mathcal{\bar{N}}_{u}^{A}$,
it follows that 
\begin{equation}
\mathcal{J}_{0}=\mathcal{N}_{A}^{u}\mathcal{\bar{N}}_{u}^{A}\qquad \text{,}%
\qquad \mathcal{J}_{a}=\mathcal{N}_{B}^{u}\left( \tau _{a}\right) _{A}^{B}%
\mathcal{\bar{N}}_{u}^{A}  \label{mab}
\end{equation}%
which read also like%
\begin{equation}
\begin{tabular}{lll}
$\mathcal{J}_{0}$ & $=$ & $\frac{1}{2}\varepsilon _{AB}\Omega _{uv}\mathcal{N%
}^{Au}\mathcal{\bar{N}}^{Bv}$ \\ 
$\mathcal{J}_{a}$ & $=$ & $\frac{1}{2}\Omega _{uv}\mathcal{N}^{u}\tau _{a}%
\mathcal{\bar{N}}^{v}$%
\end{tabular}%
\end{equation}%
By substituting $\mathcal{N}_{A}^{u}$ by its expression in terms of the
degrees of freedom of the complex hypermultiplet namely, 
\begin{equation}
\begin{tabular}{lll}
$\mathcal{N}_{A}^{u}$ & $=$ & $\left( \eta ^{i}\mathcal{\bar{F}}_{i}-g_{i}%
\bar{z}^{i}\right) K_{A}^{u}$ \\ 
$\mathcal{\bar{N}}_{u}^{A}$ & $=$ & $\left( \eta ^{i}\mathcal{F}%
_{i}-g_{i}z^{i}\right) \bar{K}_{u}^{A}$%
\end{tabular}%
\end{equation}%
with $K_{A}^{u}$ given by (\ref{gn}),%
\begin{equation}
K_{A}^{u}=\left( 
\begin{array}{c}
-\bar{k}_{u} \\ 
k^{u}%
\end{array}%
\right) \qquad \text{,}\qquad \bar{K}_{u}^{A}=\left( 
\begin{array}{c}
-k^{u} \\ 
\bar{k}_{u}%
\end{array}%
\right)
\end{equation}%
we can write down the explicit expression of $\boldsymbol{M}_{B}^{A}$ as
follows%
\begin{equation}
\boldsymbol{M}_{B}^{A}=M^{2}\text{ }\left( K_{A}^{u}\bar{K}_{u}^{B}\right)
\label{ma}
\end{equation}%
with%
\begin{equation}
M^{2}=2\left \vert \eta ^{i}\mathcal{F}_{i}-g_{i}z^{i}\right \vert ^{2}
\end{equation}%
and%
\begin{equation}
K_{A}^{u}\bar{K}_{u}^{B}=\frac{1}{2}\left( K_{A}^{u}\bar{K}_{u}^{A}\right)
\delta _{B}^{A}+\frac{1}{2}\left( K^{u}\tau _{a}\bar{K}_{u}\right) \left(
\tau ^{a}\right) _{B}^{A}
\end{equation}%
reading explicitly like%
\begin{equation}
K_{A}^{u}\bar{K}_{u}^{B}=\left( 
\begin{array}{cc}
k^{u}\bar{k}_{u} & -\left( \bar{k}_{u}\right) ^{2} \\ 
-\left( k^{u}\right) ^{2} & k^{u}\bar{k}_{u}%
\end{array}%
\right)
\end{equation}%
Then, using (\ref{xkv}), we can determine the expression of $K_{A}^{u}\bar{K}%
_{u}^{B}$ in terms of the complex scalars. We find%
\begin{equation}
\begin{tabular}{lll}
$+\bar{k}_{u}k^{u}$ & $=$ & $\left \vert q_{1}\right \vert ^{2}+\left \vert
q_{2}\right \vert ^{2}$ \\ 
$-\left( k^{u}\right) ^{2}$ & $=$ & $\left( q_{1}\right) ^{2}+\left(
q_{2}\right) ^{2}$ \\ 
$-\left( \bar{k}_{u}\right) ^{2}$ & $=$ & $\left( \bar{q}_{1}\right)
^{2}+\left( \bar{q}_{2}\right) ^{2}$%
\end{tabular}%
\end{equation}%
Adding the obtained $\boldsymbol{M}_{B}^{A}$ with the $\boldsymbol{G}%
_{B}^{A} $ contribution (\ref{ga}) coming form the gauge sector we can
compute the contribution to the $\boldsymbol{D}_{B}^{A}$ anomaly and the $%
\mathcal{V}_{\text{ADJ}}^{\mathcal{N}=2}$ scalar potential by using the
rigid Ward identity (\ref{rwi}) that we rewrite like%
\begin{equation}
\boldsymbol{G}_{B}^{A}+\boldsymbol{M}_{B}^{A}=\boldsymbol{D}_{B}^{A}+\delta
_{B}^{A}\mathcal{V}_{\text{scal}}^{\mathcal{N}=2}  \label{gd}
\end{equation}%
The contribution of $\boldsymbol{M}_{B}^{A}$ reads explicitly as $%
\boldsymbol{M}_{B}^{A}=\mathcal{J}_{0}\delta _{B}^{A}+\mathcal{J}_{a}\left(
\tau ^{a}\right) _{B}^{A}$ with 
\begin{equation}
\begin{tabular}{lll}
$\mathcal{J}_{0}$ & $=$ & $\frac{M^{2}}{2}\left( K^{u}\bar{K}_{u}\right) $
\\ 
$\mathcal{J}_{a}$ & $=$ & $\frac{M^{2}}{2}\left( K^{u}\tau _{a}\bar{K}%
_{u}\right) $%
\end{tabular}
\label{J0}
\end{equation}%
and%
\begin{equation}
\begin{tabular}{lll}
$\mathcal{J}_{0}$ & $=$ & $M^{2}\left( \left \vert q_{1}\right \vert
^{2}+\left \vert q_{2}\right \vert ^{2}\right) $ \\ 
$\mathcal{J}_{a}\tau ^{a}$ & $=$ & $M^{2}\left( 
\begin{array}{cc}
0 & \left( \bar{q}_{1}\right) ^{2}+\left( \bar{q}_{2}\right) ^{2} \\ 
\left( q_{1}\right) ^{2}+\left( q_{2}\right) ^{2} & 0%
\end{array}%
\right) $%
\end{tabular}
\label{jab}
\end{equation}%
where the effective mass $M^{2}$ is given by (\ref{ma}). We also have 
\begin{equation}
\mathcal{J}_{a}=M^{2}\left( 
\begin{array}{c}
\func{Re}\left[ \left( q_{1}\right) ^{2}+\left( q_{2}\right) ^{2}\right] \\ 
\func{Im}\left[ \left( q_{1}\right) ^{2}+\left( q_{2}\right) ^{2}\right] \\ 
0%
\end{array}%
\right)
\end{equation}%
Notice moreover that the $\eta ^{i}\mathcal{F}_{i}$ term in the expression
of $M^{2}$ generalize the hypermultiplet mass parameter in the ADJ model 
\textrm{\cite{Antoniadis:2012cg}}. This hypermultiplet mass term can be
viewed as $M^{2}=\left \vert Z\right \vert ^{2}$ describing a BPS saturation
condition \textrm{\cite{Seiberg:1994rs} with complex }$Z$ standing for the 
\textrm{central charge of the }$\mathcal{N}=2$\textrm{\ supersymmetry theory
given by}%
\begin{equation}
Z=\sqrt{2}\left( g_{i}z^{i}-\eta ^{i}\mathcal{F}_{i}\right) =\sqrt{2}%
G_{M}V^{M}
\end{equation}%
where the $g_{i}$'s are the electric charges and the $\eta ^{i}$'s their
magnetic partners. Notice as well that using (\ref{gn}) and $%
M^{2}=\left
\vert Z\right \vert ^{2}$, we can express $\mathcal{J}_{0}$ and 
$\mathcal{J}_{a}$ in terms of the symplectic gauge coupling constants $G_{M}$%
, the holomorphic section $V^{M}$ and its complex conjugate $\bar{V}^{N}$ as
follows%
\begin{equation}
\begin{tabular}{lll}
$\mathcal{J}_{0}$ & $=$ & $G_{M}G_{N}\left( K^{u}\bar{K}_{u}\right) V^{M}%
\bar{V}^{N}$ \\ 
$\mathcal{J}_{a}$ & $=$ & $G_{M}G_{N}\left( K^{u}\tau _{a}\bar{K}_{u}\right)
V^{M}\bar{V}^{N}$%
\end{tabular}%
\end{equation}%
showing that they are proportional to $\left( G_{M}\right) ^{2}$, the square
of the symplectic gauge coupling constants with coefficients like $%
g_{i}g_{j},$ $g_{i}\eta ^{j}$ and $\eta ^{i}\eta ^{j}$. From the rigid Ward
identity (\ref{rwi}), we obtain the following scalar potential%
\begin{equation}
\mathcal{V}_{\text{scal}}^{\mathcal{N}=2}=\mathcal{S}_{0}+\mathcal{J}_{0}
\label{vadj}
\end{equation}%
where $\mathcal{S}_{0}$ is as in (\ref{ss}). We also obtain the matrix
anomaly $\boldsymbol{D}_{B}^{A}$ which reads as follows 
\begin{equation}
\boldsymbol{D}_{B}^{A}=\mathcal{S}_{B}^{A}+\mathcal{J}_{B}^{A}  \label{dab}
\end{equation}%
with $\mathcal{S}_{B}^{A}$ given by eqs(\ref{bc}-\ref{ab}). Moreover, being
traceless, this $\boldsymbol{D}_{B}^{A}$ matrix can be also expressed like 
\begin{equation}
\boldsymbol{D}_{B}^{A}=\sum d_{a}\left( \tau ^{a}\right) _{B}^{A}  \label{di}
\end{equation}%
with%
\begin{equation}
d_{a}=\zeta _{a}+\alpha _{a}+\beta _{a}+\mathcal{J}_{a}  \label{da}
\end{equation}%
A non vanishing norm of the isotriplet $d_{a}$ can break partially
supersymmetry as discussed in next section.

\section{Partial breaking and ADJ model}

In this section, we study the partial breaking of $\mathcal{N}=2$
supersymmetry in the generalised APT model constructed in section 3. This
generalized APT model contains also the model of ADJ as a particular choice
of the dyonic\ FI charges and couplings.

\subsection{Partial breaking}

As in the ADFT method, the partial breaking of extended supersymmetry can
occur whenever the $\boldsymbol{D}_{B}^{A}$ matrix deformation given by eq(%
\ref{dab}) has a non vanishing VEV. This condition can be stated as 
\begin{equation}
\left \langle \boldsymbol{D}_{B}^{A}\right \rangle \neq 0\qquad
\Leftrightarrow \qquad \left \langle \zeta _{a}\right \rangle +\left \langle
\alpha _{a}\right \rangle +\left \langle \beta _{a}\right \rangle +\left
\langle \mathcal{J}_{a}\right \rangle \neq 0  \label{ad}
\end{equation}%
and it is fulfilled if one of the four isotriplets $\left \langle \zeta
_{a}\right \rangle ,$ $\left \langle \alpha _{a}\right \rangle $, $%
\left
\langle \beta _{a}\right \rangle $ and $\left \langle \mathcal{J}%
_{a}\right
\rangle $\ is different from zero. If two of these VEVs or more
are different from zero, one has to ensure that their sum is non zero. This
feature can be established by starting from the rigid Ward identity (\ref%
{rwi}) that we rewrite it as follows%
\begin{equation}
\boldsymbol{G}_{B}^{A}+\boldsymbol{M}_{B}^{A}=\delta _{B}^{A}\mathcal{V}_{%
\text{scal}}^{\mathcal{N}=2}+\boldsymbol{D}_{B}^{A}  \label{idd}
\end{equation}%
with scalar potential $\mathcal{V}_{\text{scal}}^{\mathcal{N}=2}$ given by
eqs (\ref{vadj}). By performing a similarity transformation on (\ref{idd})
by multiplying its members on right by the matrix transformation $\mathbb{U}$
and on left by its inverse $\mathbb{U}^{-1}$, we can diagonalise the
traceless matrix anomaly $\boldsymbol{D}_{B}^{A}$ like%
\begin{equation}
\boldsymbol{\tilde{D}}_{B}^{A}=%
\begin{pmatrix}
\left \vert \boldsymbol{d}\right \vert & 0 \\ 
0 & -\left \vert \boldsymbol{d}\right \vert%
\end{pmatrix}%
\qquad ,\qquad \boldsymbol{\tilde{D}}=\mathbb{U}^{-1}\boldsymbol{D}\mathbb{U}
\end{equation}%
where $\left \vert \boldsymbol{d}\right \vert $ stands for the norm of the
isovector $d^{a}$ in eq(\ref{di}). Under this transformation, the rigid Ward
identity becomes%
\begin{equation}
\boldsymbol{\tilde{G}}_{B}^{A}+\boldsymbol{\tilde{M}}_{B}^{A}=%
\begin{pmatrix}
\mathcal{V}_{\text{scal}}^{\mathcal{N}=2}+\left \vert \boldsymbol{d}\right
\vert & 0 \\ 
0 & \mathcal{V}_{\text{scal}}^{\mathcal{N}=2}-\left \vert \boldsymbol{d}%
\right \vert%
\end{pmatrix}%
\end{equation}%
So the partial breaking occurs when%
\begin{equation}
\mathcal{V}_{\text{scal}}^{\mathcal{N}=2}=\pm \left \vert \boldsymbol{d}%
\right \vert \quad ,\quad \boldsymbol{d\neq 0}  \label{pbc}
\end{equation}%
In what follows, we give an illustration of this condition and its solution
by first showing how the ADJ\ model can be recovered from this construction;
and how (\ref{pbc}) is realized in terms of the electric and magnetic FI
charges.

\subsection{Deriving the ADJ\ model}

To recover the ADJ model, we give particular values to the moment maps in (%
\ref{rm}). This is obtained by making appropriate choices of the charges
namely: $\left( 1\right) $ the values of the $\left( e_{i}^{a},m^{ia}\right) 
$ electric/magnetic FI charges; and $\left( 2\right) $ the values of the $%
\left( g_{i},\eta ^{i}\right) $ gauge coupling charges. For the case of the
electric $e_{i}^{a}$ and magnetic $m^{ia}$ FI charges, 
which can be written as follows \cite{Antoniadis:2012cg}%
\begin{equation}
\mathcal{L}_{FI}^{\left( elec\right) }=e_{i}^{a}Y^{ai}
\end{equation}%
where 
\begin{equation}
e_{i}^{a}=\frac{1}{4}\left( 
\begin{array}{c}
-\func{Im}e_{i} \\ 
\func{Re}e_{i} \\ 
\sqrt{2}\xi _{i}%
\end{array}%
\right)  \label{efiv}
\end{equation}%
are the electric FI charges vector, while the $SU\left( 2\right) _{R}$
triplets $Y^{ai}$ are the $\mathcal{N}=2$ auxiliary fields,%
\begin{equation}
Y^{ai}=\left( 
\begin{array}{c}
-2\func{Im}F^{i} \\ 
-2\func{Re}F^{i} \\ 
\sqrt{2}D^{i}%
\end{array}%
\right)  \label{n2afv}
\end{equation}%
Using the above $SO(3)\sim SU\left( 2\right) _{R}$ vector $Y^{ai}$, we can
write the magnetic FI term of \cite{Antoniadis:2012cg} as follows%
\begin{equation}
\mathcal{L}_{FI}^{\left( mag\right) }=\func{Im}\mathcal{F}_{ij}m^{ia}Y^{aj},
\notag
\end{equation}%
where the electric FI charges vector $m^{ia}$ reads explicitly as follows%
\begin{equation}
m^{ia}=\frac{1}{4}\left( 
\begin{array}{c}
m^{i} \\ 
0 \\ 
0%
\end{array}%
\right)  \label{mfiv}
\end{equation}%
Moreover, as was shown in \cite{Antoniadis:2012cg}, the complex hypermultiplet conrtibutions 
modify the above electric FI charges as follows:%
\begin{eqnarray*}
\mathcal{L}_{hyp}+\mathcal{L}_{int} &\supset &\int d^{4}\theta \left[
Q_{1}e^{-2\boldsymbol{gV}}\bar{Q}_{1}+Q_{2}e^{2\boldsymbol{gV}}\bar{Q}_{2}%
\right] +\int d^{2}\theta \left[ i\sqrt{2}\left(
\sum_{l=1}^{n}g_{l}X^{l}\right) Q_{1}Q_{2}\right] +h.c \\
&=&\mathcal{E}_{i}^{a}Y^{ia}+...
\end{eqnarray*}%
with $Y^{ia}$ are the auxiliary field vectors (\ref{n2afv}) and%
\begin{equation}
\mathcal{E}_{i}^{a}=g_{i}\digamma ^{a}  \label{cfic}
\end{equation}%
where%
\begin{equation}
\digamma ^{a}=\sqrt{2}\left( 
\begin{array}{c}
\func{Im}\left( iq_{1}q_{2}\right) \\ 
\func{Re}\left( i\bar{q}_{1}\bar{q}_{2}\right) \\ 
-\frac{1}{2}\left[ \left \vert q_{1}\right \vert ^{2}-\left \vert
q_{2}\right \vert ^{2}\right]%
\end{array}%
\right)  \label{ap}
\end{equation}%
Thus, from the above discussion we conclude that the ADJ model is obtained
by choosing the FI and coupling charges $e_{i}^{a},$ $m^{ia}$ and $\mathcal{E%
}_{i}^{a}$ in (\ref{rm}) as respectively in the equations (\ref{efiv}), (\ref%
{mfiv}) and (\ref{cfic}) while setting $\mathcal{M}^{ia}=\mathbf{0}$.

We note that thanks to the magnetic coupling, in the appendix, 
one can have the following magnetic term%
\begin{eqnarray*}
\mathcal{L}_{CS} &\supset &\int d^{2}\theta \sum_{l=1}^{n}\eta ^{i}\mathcal{F%
}_{l}\left( i\sqrt{2}Q_{1}Q_{2}\right) +hc \\
&=&\func{Im}\mathcal{F}_{ij}\mathcal{M}^{ia}Y^{aj}
\end{eqnarray*}%
with $\mathcal{M}^{ia}$ is the following magnetic coupling charge vectors%
\begin{equation}
\mathcal{M}^{ai}=\eta ^{i}\mathcal{M}^{a}=\eta ^{i}\left( 
\begin{array}{c}
\sqrt{2}\func{Im}\left( iq_{1}q_{2}\right) \\ 
\sqrt{2}\func{Re}\left( i\bar{q}_{1}\bar{q}_{2}\right) \\ 
0%
\end{array}%
\right)
\end{equation}%
By substituting the above choices back into eq(\ref{vadj}), the effective
scalar potential $\mathcal{V}_{scal}=\mathcal{S}_{0}+\mathcal{J}_{0}$ is
given by%
\begin{equation}
\begin{tabular}{lll}
$\mathcal{V}_{scal}$ & $=$ & $\left. 2\left \vert \eta ^{i}\mathcal{F}%
_{i}-g_{i}z^{i}\right \vert ^{2}\right \vert _{\eta =0}\left( \left \vert
q_{1}\right \vert ^{2}+\left \vert q_{2}\right \vert ^{2}\right) $ \\ 
&  & $+\frac{1}{16}\mathcal{G}^{i\bar{j}}\left( e_{i}+im^{k}\mathcal{F}%
_{ik}-4i\sqrt{2}g_{i}\right) \left( e_{\bar{j}}-im^{l}\mathcal{\bar{F}}_{%
\bar{j}l}+4i\sqrt{2}g_{\bar{j}}\bar{q}_{1}\bar{q}_{2}\right) $ \\ 
&  & $+\frac{1}{8}\mathcal{G}^{i\bar{j}}\left[ \xi _{i}-2g_{i}\left( \left
\vert q_{1}\right \vert ^{2}-\left \vert q_{2}\right \vert ^{2}\right) %
\right] \left[ \xi _{\bar{j}}-2g_{\bar{j}}\left( \left \vert q_{1}\right
\vert ^{2}-\left \vert q_{2}\right \vert ^{2}\right) \right] $%
\end{tabular}%
\end{equation}%
Comparing this expression with the ADJ model, we learn that $\mathcal{V}%
_{scal}$ is precisely the scalar potential $\mathcal{V}_{\text{ADJ}}^{%
\mathcal{N}=2}$ given by eq(3.1) of \cite{Antoniadis:2012cg} where the scalar field $z_{k}$ is
shifted by a constant like%
\begin{equation}
z_{k}\rightarrow z_{k}-\frac{im}{\sqrt{2}}\frac{g_{k}}{g^{2}}\qquad ,\qquad
g^{2}=\sum_{i=1}^{2}g_{i}
\end{equation}%
These shifts of $z_{k}$ correspond to replace the electric central charge $%
Z=i\sqrt{2}\left. \left( \eta ^{i}\mathcal{F}_{i}-g_{i}z^{i}\right)
\right
\vert _{\eta =0}$ of our construction by 
\begin{equation}
Z=m+ig_{i}z^{i}\sqrt{2}
\end{equation}%
Notice that VEV of $\mathcal{V}_{scal}$ depends on the coupling constants
and the VEVs of the scalar fields of the theory namely the complex doublet $%
\left( q_{1},q_{2}\right) $ and the symplectic $\left( z^{i},\mathcal{F}%
_{i}\right) $. If for instance the condition $\frac{\partial \mathcal{V}%
_{scal}}{\partial q^{u}}=0$ is solved by $\left \langle q_{u}\right \rangle
=0$, we have%
\begin{equation}
\left \langle \mathcal{V}_{scal}\right \rangle =\frac{1}{16}\mathcal{G}^{i%
\bar{j}}\left( e_{i}+im^{k}\left \langle \mathcal{F}_{ik}\right \rangle
\right) \left( e_{\bar{j}}-im^{l}\left \langle \mathcal{\bar{F}}_{\bar{j}%
l}\right \rangle \right) +\frac{1}{8}\mathcal{G}^{i\bar{j}}\xi _{i}\xi _{%
\bar{j}}\geq 0  \label{qvs}
\end{equation}%
with $\left \langle \mathcal{F}_{ik}\right \rangle =\mathcal{F}_{ik}\left[
\left \langle z\right \rangle \right] $ where the $\left \langle
z^{i}\right
\rangle $'s solve the condition $\frac{\partial \mathcal{V}%
_{scal}}{\partial z^{i}}=0$. For these values the anomaly vector (\ref{da})
reads as%
\begin{equation}
\left \langle d_{c}\right \rangle =2\varepsilon
_{abc}m^{ai}e_{i}^{b}=2\left( \vec{m}^{i}\wedge \vec{e}_{i}\right) _{c}
\end{equation}%
where $e_{i}^{a}$ and $m^{bi}$ are the electric and magnetic charges eqs(\ref%
{efiv}, \ref{mfiv}). Thus, the norm $\left \langle |\boldsymbol{d}%
|\right
\rangle $ characterising the partial breaking of $\mathcal{N}=2$
supersymmetry reads as follows%
\begin{equation}
\left \langle |\boldsymbol{d}|\right \rangle =\frac{1}{8}\sqrt{%
m^{i}m^{j}\left( \func{Re}e_{i}\func{Re}e_{j}+2\xi _{i}\xi _{j}\right) }
\label{qd}
\end{equation}%
If we further choose the complex electric FI charges $e_{i}$ to be pure
imaginary, the VEVs of the scalar potential (\ref{qvs}) and the norm (\ref%
{qd}) become%
\begin{eqnarray}
\left \langle \mathcal{V}_{scal}\right \rangle &=&\frac{1}{16}\mathcal{G}^{i%
\bar{j}}\left( \func{Im}e_{i}+m^{k}\mathcal{F}_{ik}\right) \left( \func{Im}%
e_{j}+m^{l}\mathcal{\bar{F}}_{\bar{j}l}\right) +\frac{1}{8}\mathcal{G}^{i%
\bar{j}}\xi _{i}\xi _{j}  \notag \\
\left \langle |\boldsymbol{d}|\right \rangle &=&\frac{1}{4\sqrt{2}}m^{i}\xi
_{i}
\end{eqnarray}%
Hence, the partial breaking condition (\ref{pbc}) becomes%
\begin{equation}
\mathcal{G}^{i\bar{j}}\left( \func{Im}e_{i}+m^{k}\mathcal{F}_{ik}\right)
\left( \func{Im}e_{j}+m^{l}\mathcal{\bar{F}}_{\bar{j}l}\right) +2\mathcal{G}%
^{i\bar{j}}\xi _{i}\xi _{\bar{j}}=\pm 2\sqrt{2}m^{i}\xi _{i}
\end{equation}%
which can be rewritten as follows%
\begin{equation}
\left( \func{Im}e_{\bar{j}}+m^{k}\func{Re}\mathcal{F}_{k\bar{j}}\right)
^{2}+\left( \mathcal{G}_{i\bar{j}}m^{_{\bar{j}}}\mp \sqrt{2}\xi _{i}\right)
^{2}=0
\end{equation}%
and so has the following solutions%
\begin{equation}
\xi _{i}=\pm \frac{1}{\sqrt{2}}\mathcal{G}_{i\bar{j}}m^{_{\bar{j}}}\quad
,\quad \func{Im}e_{\bar{j}}=-m^{k}\func{Re}\mathcal{F}_{k\bar{j}}
\end{equation}%
which coincides with the partial breaking condition given in eq(4.9) of the \textrm{ADJ
model} \cite{Antoniadis:2012cg}.

\section{Conclusion}

In this paper we have studied the partial breaking of rigid $\mathcal{N}=2$
supersymmetric gauge theory of $n$ vector multiplets coupled to complex
hypermultiplets by using the ADFT method given in \cite{Andrianopoli:2015wqa}%
. To that purpose, we have first reviewed the ADFT method where partial
breaking of rigid supersymmetry is induced by an anomalous isotriplet vector 
$\zeta _{a}$ originating from the hidden sector in the rigid limit of gauged 
$\mathcal{N}=2$ supergravity. This isovector has the form $\zeta _{a}\sim
\varepsilon _{abc}m^{a}e^{b}$ with the isotriplet $e^{a}$ and $m^{a}$
standing for the electric and magnetic Fayet- Iliopoulos terms. In our
construction, which may be viewed as a generalisation of ADFT method by adding
complex hypermatter coupled to gauge degrees of freedom, we showed that the
anomaly $\zeta _{a}$ gets three extra contributions given by (\ref{ad})
namely $\left \langle d_{a}\right \rangle =\left \langle \zeta
_{a}\right
\rangle +\left \langle \alpha _{a}\right \rangle +\left \langle
\beta _{a}\right \rangle +\left \langle \mathcal{J}_{a}\right \rangle $. The 
$\zeta _{a}$ is as in ADFT method, and the three other contributions $%
\left
\langle \alpha _{a}\right \rangle ,$ $\left \langle \beta
_{a}\right
\rangle $ and $\left \langle \mathcal{J}_{a}\right \rangle $ are
induced by the presence of complex hypermatter. The $\left \langle \alpha
_{a}\right
\rangle $ is induced by the coupling between FI constant and the
electric-magnetic coupling charges. The $\left \langle \beta
_{a}\right
\rangle $ is due to the coupling between electric and magnetic
charges and $\left \langle \mathcal{J}_{a}\right \rangle $ to local dyonic
mass. Non zero contribution of these anomalies are dependent on non-zero
VEVS of the hypermatter fields $\left \langle q^{u}\right \rangle $
determined by minimizing the scalar potential. Their expressions have been
explicitly studied in subsection 3.2. \newline
In the extension of the ADFT\ method developed in this paper, we also gave
the rigid Ward identity and the induced scalar potential of as well as the
general condition for partial breaking. By choosing a particular values of
the components of our generalized moment maps, we derived as well the scalar
potential of the ADJ model and their partial breaking condition. It would be
interesting to obtain the extended APT method, studied in the present paper,
as an observable sector in the rigid limit of $\mathcal{N}=2$ gauged
supergravity coupled to complex hypermatter. Progress in this direction will
be reported in a future occasion.

\appendix

\section{Dyonic couplings}

First, we recall that in $\mathcal N=2$ supersymmetry one can distinguish two multiplets:

\begin{itemize}
	\item Vector multiplet: which has the following $\tilde{\theta}$- expansion 
	\begin{equation}
	\mathcal{W}\left( x,\theta ,\tilde{\theta}\right) =X\left( x,\theta
	\right) +i\sqrt{2}\tilde{\theta}\boldsymbol{W}\left( x,\theta \right) +%
	\tilde{\theta}^{2}\left( -\frac{1}{4}\bar{D}^{2}\bar{X}^{i}\right)
	\label{wi}
	\end{equation}%
	where $X\left( x,\theta \right) $ and $\boldsymbol{W}\left( x,\theta
	\right) $ are $\mathcal{N}=1$ chiral superfields where the fermionic $%
	\boldsymbol{W}\left( x,\theta \right) $'s are precisely the $\mathcal{N}%
	=1$ superfield strengths living inside the $\mathcal{N}=2$ ones.
	
	\item Hypermultiplet: which can be described by two $\mathcal{N}=1$ chiral superfields $%
	Q^{u}=\left( Q^{1},Q^{2}\right) $ having opposite gauge charges and $\theta $%
	- expansions as follows%
	\begin{equation}
	Q^{u}=q^{u}+\sqrt{2}\theta \chi ^{u}+\theta \theta G^{u}
	\end{equation}%
	where $G^{u}$ are auxiliary fields. 
\end{itemize}

The aim of this appendix is to give the extension of ADJ model by allowing dyonic
gauge couplings. To that purpose, we are interested only in the non-kinetic terms of the action, namely the quadratic complex mass term \cite{Antoniadis:2012cg}
\begin{equation}
\mathcal{L}_{mass}=m\int d^{2}\theta Q_{1}Q_{2}+h.c  \label{m}
\end{equation}%
where $Q_1$ and $Q_2$ are the two $\mathcal N=1$ chiral superfields constituting the $\mathcal N=2$ hypermultiplet, and the couplings to the gauge to the $X^{i}$ chiral superfields, of the $\mathcal N=2$ vector multiplet, given by the following tri-superfield interactions, 
\begin{equation}
\mathcal{L}_{int}=\int d^{2}\theta \left[ i\sqrt{2}\left(
\sum_{l=1}^{n}g_{l}X^{l}\right) Q_{1}Q_{2}\right] +h.c  \label{lint},
\end{equation}
where $X^l$, with $l=1,...,n_v$ are the first components of the $\mathcal N=2$ vector superfields $\mathcal{W}^{l}$. 

We will start from the Lagrangian density (\ref%
{lint})\ and show that it has an interpretation in terms of $\mathcal{N}=2$
Chern-Simon interaction using the dual tensorial description of the
hypermultiplet $\left( Q_{1},Q_{2}\right) $. Then, we turn to study the $%
\mathcal{N}=2$ dyonic gauge invariant couplings

\subsection{$\mathcal{N}=2$ Chern-Simons action} 

In this description, the $\mathcal{L}_{int}$ can be derived by starting from
the $\mathcal{N}=2$ chiral superspace action%
\begin{equation}
\mathcal{L}_{CS}=-2i\int d^{2}\theta d^{2}\tilde{\theta}\left(
\sum_{l=1}^{n}g_{l}\mathcal{W}^{l}\right) \mathcal{T}^{N=2}+hc
\label{cs}
\end{equation}%
\ where $\mathcal{T}^{N=2}$ is the $\mathcal{N}=2$ tensor superfield with
expansion as follows%
\begin{equation}
\mathcal{T}^{N=2}=Y\left( x,\theta \right) +\sqrt{2}\tilde{\theta}\Upsilon
\left( x,\theta \right) -\tilde{\theta}^{2}\left( \frac{i}{2}\Phi +\frac{1}{4%
}\bar{D}^{2}\bar{Y}\right)  \label{te}
\end{equation}%
In this expansion $Y\left( x,\theta \right) $ and $\Phi $ are bosonic $%
\mathcal{N}=1$ \textrm{chiral superfields} while $\Upsilon _{\alpha }$ is a
spinor $\mathcal{N}=1$ chiral superfield related to the $\mathcal{N}=1$
linear superfield $L$ like%
\begin{equation}
L=D\Upsilon +\bar{D}\bar{\Upsilon}\qquad ,\qquad DDL=\bar{D}\bar{D}L=0
\end{equation}%
The degrees of freedom are carried by the superfields $\Phi $ and $L$ as the
Y can be gauged out. Indeed, notice that in the above $\tilde{\theta}$-
expansion of $\mathcal{T}^{N=2}$, the spinor superfield $\Upsilon _{\alpha }$
is defined up to the following gauge transformation%
\begin{equation}
\delta _{gauge}\Upsilon _{\alpha }=-iW_{\alpha }^{\prime }\qquad ,\qquad
W_{\alpha }^{\prime }=-\frac{1}{4}\bar{D}\bar{D}D_{\alpha }\Delta ^{\prime }
\label{dlg1}
\end{equation}%
where $\Delta ^{\prime }$ is a general $\mathcal{N}=1$ real superfield and $%
W_{\alpha }^{\prime }$ is its field strength. The $\mathcal{N}=2$ superspace
version of (\ref{dlg1}) is given by%
\begin{equation}
\delta _{gauge}\mathcal{T}^{N=2}=-\mathcal{W}^{\prime }  \label{n2gt}
\end{equation}%
where $\mathcal{W}^{\prime }$ is a $\mathcal{N}=2$ \textrm{vector superfield
strength}. By $\tilde{\theta}$- expansion of both sides of this
transformation, we obtain%
\begin{eqnarray}
\delta _{gauge}Y &=&\frac{1}{2}\bar{D}\bar{D}\Delta ^{\prime }  \notag \\
\delta _{gauge}\Upsilon _{\alpha } &=&-iW_{\alpha }^{\prime } \\
\delta _{gauge}\Phi &=&0  \notag
\end{eqnarray}%
Notice that the gauge transformation$\ $(\ref{n2gt}) is just the $\mathcal{N}%
=2$ \textrm{superspace version of the gauge transformation} $\delta
_{gauge}b^{\rho \sigma }=\partial ^{\rho }\Lambda ^{\sigma }$, where $%
b^{\rho \sigma }$ is a $2$-form field whose field strength $3$-form is
unchanged under this gauge transformation \cite{Ambrosetti:2009za}. The
gauge transformation (\ref{n2gt}) allows us to eliminate the superfield$\ Y$
but one can instead choose a gauge in wich is has it has only a non
vanishing imaginary part of the auxiliary field,%
\begin{equation}
Y^{gauged}=\frac{i}{4!}\theta ^{2}\varepsilon _{\mu \nu \rho \sigma }F^{\mu
\nu \rho \sigma }
\end{equation}%
where $F^{\mu \nu \rho \sigma }$ is a 4-form. Substituting (\ref{wi}) and (\ref{te}) back into (\ref{cs}), we obtain the following CS lagrangian density%
\begin{equation}
\mathcal{L}_{CS}=-\int d^{2}\theta \left[ \sum_{l=1}^{n}g_{l}\left(
X^{l}\Phi \right) +\sum_{l=1}^{n}g_{l}\left( \boldsymbol{W}_{\alpha
}^{l}\Upsilon ^{\alpha }\right) +i\sum_{l=1}^{n}g_{l}\left( m^{l}Y\right) %
\right] +hc  \label{tt}
\end{equation}%
Comparing this action with (\ref{m}-\ref{lint}), we learn that the gauge
invariant $\Phi $ can be realised as the product of two chiral superchamps
as follows%
\begin{equation}
\Phi =i\sqrt{2}Q_{1}Q_{2}
\end{equation}%
The mass term $mQ_{1}Q_{2}$ can be generated by using thegauge invariant
shift $X^{l}$ $\rightarrow X^{l}-\frac{i}{\sqrt{2}}\mu ^{l}$ and setting $%
m=\sum_{l=1}^{n}g_{l}\mu ^{l}$.

\subsection{$\mathcal{N}=2$ dyonic gauge couplings}

Here, we want to comment on a property of the $\mathcal{N}=2$ superfield
lagrangian density (\ref{cs}). This superdensity has a remarkable dependence
on the $\mathcal{N}=2$ prepotentials namely on the quantity 
\begin{equation}
\Theta =\left( \sum_{l=1}^{n}g_{l}\mathcal{W}^{l}\right) \mathcal{T}^{N=2}
\label{x}
\end{equation}%
which leads, after integration with respect to $\tilde{\theta},$ to%
\begin{equation}
\int d^{2}\tilde{\theta}\Theta =\sum_{l=1}^{n}\left( g_{l}X^{l}\right) \Phi
+\sum_{l=1}^{n}\left( g_{l}\boldsymbol{W}_{\alpha }^{l}\right) \Upsilon
^{\alpha }
\end{equation}%
As the $\mathcal{N}=1$ chiral superfield combination $gX=%
\sum_{l=1}^{n}g_{l}X^{l}$ is just a half part of a symplectic invariant $%
\mathcal{N}=1$ chiral superfield namely%
\begin{equation}
\sum_{l=1}^{n}\left( g_{l}X^{l}-i\eta ^{l}\frac{\partial \mathcal{F}}{%
\partial X^{l}}\right) =G^{M}\mathcal{C}_{MN}V^{M}
\end{equation}%
with 
\begin{equation}
G^{M}=\left( 
\begin{array}{c}
i\eta ^{l} \\ 
g_{l}%
\end{array}%
\right) \qquad ,\qquad V^{M}=\left( 
\begin{array}{c}
X^{l} \\ 
\frac{\partial \mathcal{F}}{\partial X^{l}}%
\end{array}%
\right)
\end{equation}%
where $\eta ^{l}$ are real parameters. One may think about (\ref{x}) as just
a part of the following symplectic invariant quantity 
\begin{equation}
\Xi =\left( \sum_{l=1}^{n}g_{l}\mathcal{W}^{l}-i\eta ^{l}\frac{\partial 
\mathcal{F}}{\partial \mathcal{W}^{l}}\right) \mathcal{T}^{N=2}
\end{equation}%
These relations suggest that the $\mathcal{N}=2$ Chern-Simons action (\ref%
{cs}) can be made symplectic invariant as follows%
\begin{equation}
\mathcal{L}_{CS}=\int d^{2}\theta d^{2}\tilde{\theta}\sum_{l=1}^{n}\left(
g_{l}\mathcal{W}^{l}-i\eta ^{l}\frac{\partial \mathcal{F}}{\partial \mathcal{%
W}^{l}}\right) \mathcal{T}^{N=2}+hc  \label{sc}
\end{equation}%
which, by integration with respect to $\tilde{\theta}$, leads to%
\begin{equation}
\mathcal{L}_{CS}=\int d^{2}\theta \sum_{l=1}^{n}\left( g_{l}X^{l}-i\eta ^{l}%
\mathcal{F}_{l}\right) \Phi +\sum_{l=1}^{n}\left( g_{l}-i\eta ^{j}\mathcal{F}%
_{jl}\right) \boldsymbol{W}_{\alpha }^{l}\Upsilon ^{\alpha }
\end{equation}%
with%
\begin{equation}
\mathcal{F}_{i}=\frac{\partial \mathcal{F}\left( X\right) }{\partial X^{i}}%
\qquad ,\qquad \mathcal{F}_{ij}=\frac{\partial ^{2}\mathcal{F}\left(
X\right) }{\partial X^{i}\partial X^{j}}
\end{equation}%
\begin{equation*}
\end{equation*}


\end{document}